# Megahertz single-particle imaging at the European XFEL

Egor Sobolev[1][†], Serguey Zolotarev[2][†], Klaus Giewekemeyer[3], Johan Bielecki[3], Kenta Okamoto[4], Hemanth K. N. Reddy[4], Jakob Andreasson[5], Kartik Ayyer[6,7], Imrich Barak[8], Sadia Bari[9], Anton Barty[6], Richard Bean[3], Sergey Bobkov[2], Henry N. Chapman[6,13,24,], Grzegorz Chojnowski[1], Benedikt J. Daurer[4,10], Katerina Dörner[3], Tomas Ekeberg[4], Leonie Flückiger[11], Oxana Galzitskaya[25,26], Luca Gelisio[9], Steffen Hauf[3], Brenda G. Hogue[12], Daniel A. Horke[6,13], Ahmad Hosseinizadeh[14], Vyacheslav Ilyin[2], Chulho Jung[15], Chan Kim[3], Yoonhee Kim[3], Richard A. Kirian[16], Henry Kirkwood[3], Olena Kulyk[5], Romain Letrun[3], Duane Loh[10], Marc Messerschmidt[17], Kerstin Mühlig[4], Abbas Ourmazd[14], Natascha Raab[3], Andrei V. Rode[18], Max Rose[9], Adam Round[3], Takushi Sato[3], Robin Schubert[3], Peter Schwander[14], Jonas A. Sellberg[19], Marcin Sikorski[3], Alessandro Silenzi[3], Changyong Song[15], John C. H. Spence[16], Stephan Stern[3], Jolanta Sztuk-Dambietz[3], Anthon Teslyuk[2], Nicusor Timneanu[20], Martin Trebbin[21], Charlotte Uetrecht[3,22], Britta Weinhausen[3], Garth J. Williams[23], P Lourdu Xavier[6,7], Chen Xu[3], Ivan Vartanyants[9], Victor S. Lamzin[1], Adrian Mancuso[3], Filipe R. N. C. Maia[4,27] *

[†] These authors contributed equally to the work.
* Corresponding author (filipe@xray.bmc.uu.se).

**Affiliations**
1) European Molecular Biology Laboratory, c/o DESY, Notkestrasse 85, 22607 Hamburg, Germany.
2) NRC "Kurchatov Institute" Pl. Akademika Kurchatova, 1, 123098 Moscow, Russia.
3) European XFEL GmbH, Holzkoppel 4, 22869 Schenefeld, Germany.
4) Laboratory of Molecular Biophysics, Department of Cell and Molecular Biology, Uppsala University, Husargatan 3 (Box 596), SE-75124 Uppsala, Sweden.
5) Institute of Physics, ELI Beamlines, Academy of Sciences of the Czech Republic, Na Slovance 2, CZ-18221 Prague, Czech Republic.
6) Center for Free-Electron Laser Science, Deutsches Elektronen-Synchrotron DESY, Notkestrasse 85, 22607 Hamburg, Germany.
7) Max Planck Institute for the Structure and Dynamics of Matter, Luruper Chaussee 149, 22761 Hamburg, Germany.
8) Institute of Molecular Biology, SAS, Dubravska cesta 21, 845 51 Bratislava, Slovakia.
9) Deutsches Elektronen-Synchrotron DESY, Notkestrasse 85, 22607 Hamburg, Germany.
10) Centre for BioImaging Sciences, Department of Biological Sciences, National University of Singapore, 14 Science Drive 4, Singapore 117557, Singapore.
11) ARC Centre of Advanced Molecular Imaging, Department of Chemistry and Physics, La Trobe University, Melbourne, 3086, Australia.
12) Arizona State University, School of Life Sciences (SOLS), Tempe, AZ 85287, USA
13) The Hamburg Center for Ultrafast Imaging, Universität Hamburg, Luruper Chaussee 149, 22761 Hamburg, Germany.
14) Department of Physics, University of Wisconsin Milwaukee, KIRC, 3135 N. Maryland Ave, Milwaukee, Wisconsin 53211, USA.
15) Department of Physics, Pohang University of Science and Technology, Pohang 37673, Korea.
16) Department of Physics, Arizona State University, Tempe, Arizona 85287, USA.
17) Biodesign Institute at Arizona State University, Tempe, Arizona 85287-5001, USA.
18) Laser Physics Centre, Research School of Physics, Australian National University, Canberra, ACT 2601, Australia.
19) Biomedical and X-ray Physics, Department of Applied Physics, AlbaNova University Center, KTH Royal Institute of Technology, SE-10691 Stockholm, Sweden.
20) Department of Physics and Astronomy, Uppsala University, Box 516, SE-75120 Uppsala, Sweden.
21) State University of New York at Buffalo, Department of Chemistry, Natural Sciences Complex, Room 760, Buffalo, New York 14260-3000, USA.
22) Heinrich Pette Institute, Leibniz Institute for Experimental Virology, Martinistrasse 52, Hamburg 20251, Germany.
23) NSLS-II, Brookhaven National Laboratory, UPTON, NY, 11772 USA.
24) Department of Physics, Universität Hamburg, Luruper Chaussee 149, 22761 Hamburg, Germany.
25) Laboratory of Bioinformatics and Proteomics, Institute of Protein Research, Russian Academy of Sciences, 142290 Pushchino, Moscow Region, Russia.

26) Institute of Theoretical and Experimental Biophysics, Russian Academy of Sciences, 142290 Pushchino, Moscow Region, Russia.
27) NERSC, Lawrence Berkeley National Laboratory, Berkeley, CA 94720, USA.

The emergence of high repetition-rate X-ray free-electron lasers (XFELs) powered by superconducting accelerator technology enables the measurement of significantly more experimental data per day than was previously possible. The European XFEL will soon provide 27,000 pulses per second, more than two orders of magnitude more than any other XFEL. The increased pulse rate is a key enabling factor for single-particle X-ray diffractive imaging, which relies on averaging the weak diffraction signal from single biological particles. Taking full advantage of this new capability requires that all experimental steps, from sample preparation and delivery to the acquisition of diffraction patterns, are compatible with the increased pulse repetition rate. Here, we show that single-particle imaging can be performed using X-ray pulses at megahertz repetition rates. The results obtained pave the way towards exploiting high repetition-rate X-ray free-electron lasers for single-particle imaging at their full repetition rate.

The ability of extremely intense and brief femtosecond XFEL pulses to outrun radiation damage avoids the need to freeze (and thus immobilize) biological samples to minimize damage, as required in conventional protein crystallography (Neutze et al. 2000) or cryogenic electron microscopy (cryoEM). For single particles, this enables the study of protein dynamics under near-physiological conditions at room temperature. The principle of outrunning damage by collecting diffraction data before the onset of the damaging photoelectron cascade was first established experimentally at the FLASH facility in 2006 (Chapman et al. 2006) and is now routine in serial femtosecond crystallography (Chapman et al. 2011, Aquila et al. 2012, Boutet et al. 2012). Since the first aerosol single-particle imaging experiments at FLASH (Bogan et al. 2008), the method of flash X-ray imaging (FXI) has been applied to image living cells (van der Schot et al. 2015), cell organelles (Hantke et al. 2014) and viruses (Munke et al. 2016; Reddy et al. 2017), in particular, the giant Mimivirus in 2D projections (Seibert et al. 2011), as well as in full 3D (Ekeberg et al. 2015). Despite continual improvements in reconstruction algorithms, the number of reconstructed resolution elements across the sample remains at about a dozen voxels (Kurta et al. 2017; Rose et al. 2018; Lundholm et al. 2018). The main reasons for this limitation are the large dynamic range spanned by the diffracted intensities, going beyond the technical limits of current detector technology, as well as the weakness of the diffraction signal and the shot-to-shot variations in imaging conditions due to lateral distance between the sample and the X-ray focus (the impact parameter), background scattering, and detector response. Averaging over a very large number of single-particle snapshots is required to obtain sufficient information at high-resolution regions in diffraction space. This is necessary even for strongly scattering samples. Until now, this has been hampered by the low hit probabilities and the relatively low 120 Hz pulse repetition rate at XFEL facilities available to date.

The European XFEL (EuXFEL) introduces an era of high-intensity, high-repetition-rate, and high data-rate XFELs by taking advantage of a superconducting linear accelerator (Altarelli & Mancuso, 2014). The high repetition rate poses new challenges for sample injectors and X-ray detectors. Whenever the XFEL pulse hits a sample, it rapidly transforms it into a plasma. To fully exploit the high repetition rate, this plasma must not interfere with the delivery of the next particle, thereby ensuring that different pulses correspond to independent measurements from undamaged, intact objects. For serial crystallography at the EuXFEL, this has recently been shown to be possible (Wiedorn et al. 2018; Grünbein et al. 2018, Yefanov et al. 2019).

The first single-particle experiments at the EuXFEL were performed in December 2017 using the Single Particles, Clusters, and Biomolecules & Serial Femtosecond Crystallography (SPB/SFX) instrument (Mancuso et al. 2019) with microfocus optics. The main goal of the experiment was to demonstrate single-particle imaging at the high intra--bunch repetition rate of the EuXFEL with the Adaptive Gain Integrated Pixel Detector (AGIPD) (Allahgholi, A. et al 2019). In this article, we present the results of this experiment. We start by characterizing the background inherent to the instrument, which is a critical parameter for determining the maximum achievable resolution, as well as the signal-to-noise ratio of the recorded patterns, instrumental stability, and the incident photon flux. We then size the particles corresponding to the patterns recorded while injecting viruses into the beam, confirming that a substantial fraction of the patterns corresponded to the expected particle size. Finally, we searched for any correlation or dependence among diffraction patterns obtained from the same pulse train. Overall, we show that single-particle imaging experiments can be performed at the megahertz intra-bunch repetition rate of the EuXFEL.

# RESULTS

## Overview of data collection

The experiment (p2013), was performed over five 12-hour shifts in December 2017. Data were recorded during 376 experimental runs. Each run contained 30,000 pulses, corresponding to one thousand bunch trains, with each containing 30 pulses. In total, 11 255 800 frames were recorded with the MHz camera AGIPD, out of which 557 675 patterns were identified as hits or diffraction patterns from the target samples. The overall statistics of the measured data are summarized in Table 1.

**Table 1.** The summary of measurements broken down by samples and shifts.

| Sample | Shift | # of runs | # of images | # of hits | hit ratio, % |
|---|---|---|---|---|---|
| Iridium(III) chloride | 2 | 36 | 1 162 950 | 15 348 | 1.3 |
| | 3 | 80 | 2 483 850 | 127 747 | 5.1 |
| | 4 | 21 | 630 000 | 165 620 | 26.3 |
| | 5 | 19 | 570 000 | 40 724 | 7.1 |
| | **Total** | **156** | **4 846 800** | **349 439** | **7.2** |
| Caesium iodide | Total (4) | 9 | 270 000 | 58 256 | 21.6 |
| Mimivirus | 3 | 54 | 1 620 000 | 4 140 | 0.26 |
| | 4 | 100 | 3 000 000 | 132 150 | 4.4 |
| | **Total** | **154** | **4 620 000** | **136 290** | **3.0** |
| Melbourne virus | 4 | 14 | 420 000 | 11 416 | 2.7 |
| | 5 | 5 | 150 000 | 2 274 | 1.5 |
| | **Total** | **19** | **570 000** | **13 690** | **2.4** |

Examples of scattering from IrCl spheres and Mimivirus are shown in Fig 1.

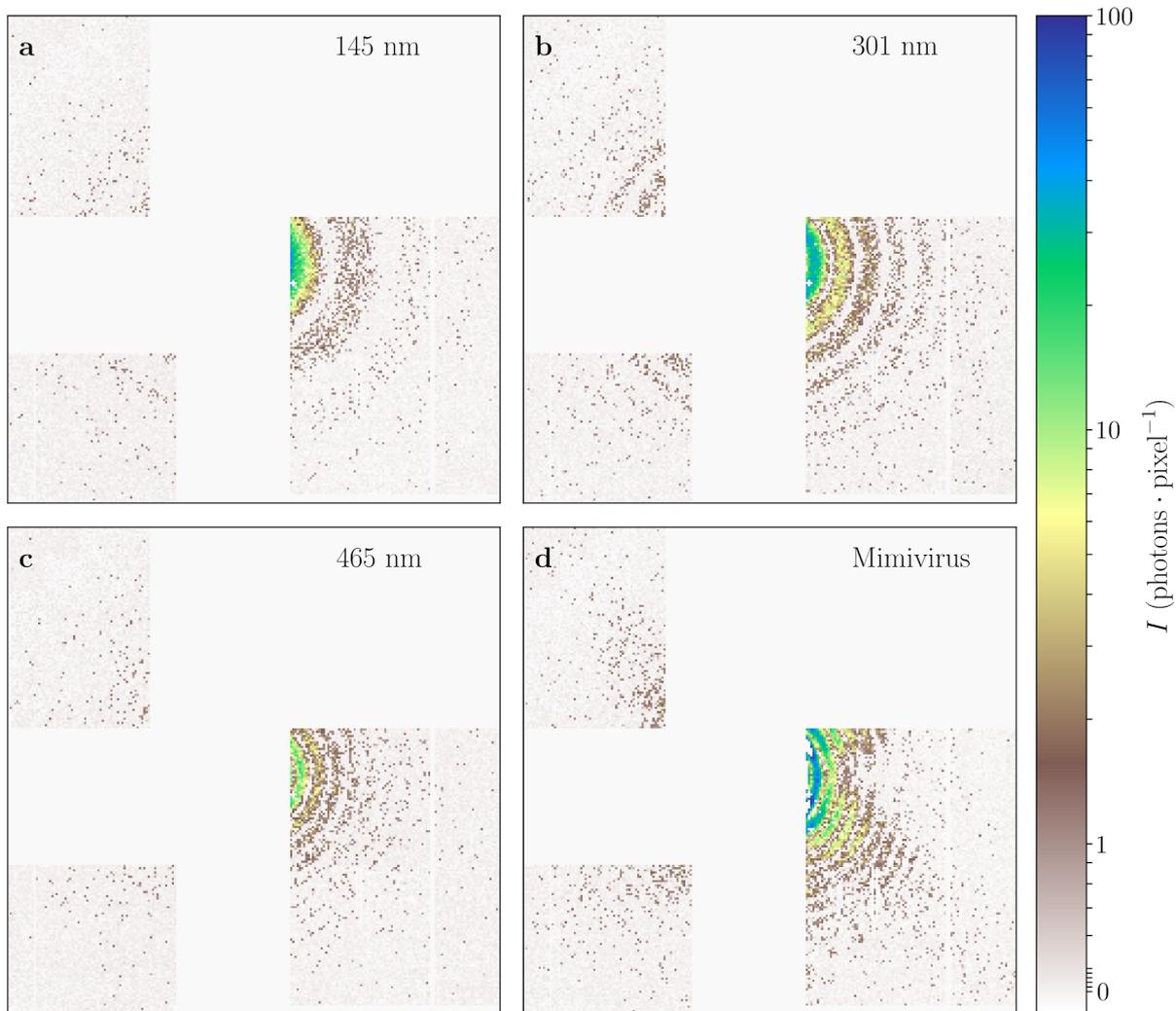

**Figure 1. a,b,c** Examples of scattering from IrCl spheres of 145 nm, 301 nm, and 465 nm diameter, respectively. **d** Scattering from Mimivirus. The edge resolution of the patterns shown is 36.8 nm.

A heavy-metal salt solution was used to align the beam and the injector. When a salt solution is aerosolized and focused by a gas dynamic virtual nozzle (GDVN) (see Methods), it forms a single-file stream of droplets. Water quickly evaporates from the droplets in a vacuum, resulting in amorphous salt spheres. In aerosol imaging experiments, a salt solution is convenient for detecting the X-ray beam since each droplet gives rise to a salt particle, thus leading to high hit-rate. This contrasts with colloidal particles dispersed in a volatile medium, where many droplets may not contain particles or form any upon injection, leading to low a hit-rate.

Diffraction from these spheres was simulated to determine the effect of experimental parameters such as the incident flow rate, particle size, and alignment on the diffraction patterns. A scattering model for spherical particles (Starodub et al, 2005) was fitted to the diffraction patterns for the iridium(III) chloride (IrCl) samples (see Fig. 1a,b,c) captured in the third and fourth shifts, as described in Methods. We assumed that the density of amorphous IrCl particles formed in vacuum was close to its solid-state density of 5.3 g/cm$^3$. Also, we assumed that, on average, each IrCl molecule is hydrated by one water molecule, resulting in a molar mass of 316.6 g/mol and a scattering factor of 133.6 electrons. Particle sizes and incident beam fluences were obtained as described in the Methods section and are shown in Fig. 2.

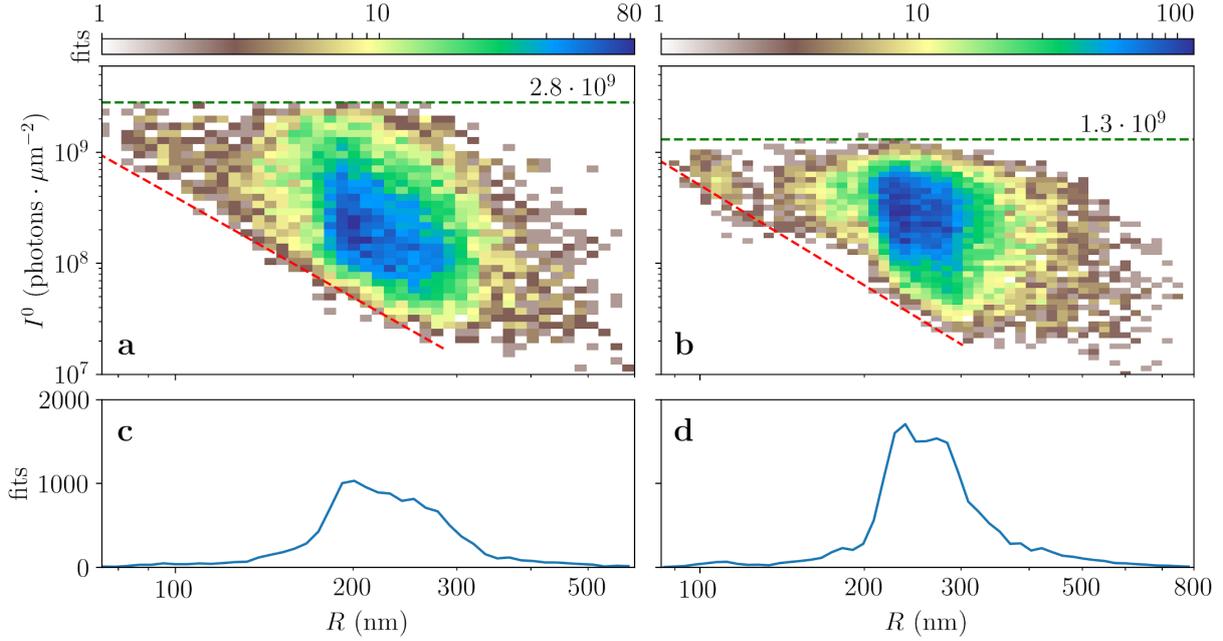

**Figure 2.** Distributions of the reconstructed parameters of scattering from spherical particles formed by IrCl. Distribution of incident photon fluences over particle diameters, shown as a 2D histogram, in the third (**a**) and fourth (**b**) shifts.

The two-dimensional distributions of particle sizes indicate that the particle size ranges from 80 nm to 800 nm in diameter (Fig. 2) and show an upper limit of the fluence of the incident photons, independent of particle size (see Fig. 2a,b, green dashed line). This limit is the value of the fluence at the focus of the beam ($I_m$), where it reaches a maximum. The lack of events in the upper-right corner of the distribution results from the small number of large particles in the measured set. Thus, we can only approximately estimate the upper limit of the flux at about 2.8×10⁹ photons/μm² during the third shift, and about 1.3×10⁹ photons/μm² during the fourth shift.

The lower fluence limit (Fig. 2a,b, red dashed line) depends on the particle size and corresponds to the sensitivity limit ($I_s$) below which it was impossible to fit a spherical scattering model. The slope of the lower bound is –3 on the log-scale, matching the scaling of the signal for a given particle volume

$$I_s = I_m(R_0^3/R^3).$$

The line showing the limit of sensitivity crosses the line for the upper limit of the flux $I_m$ at a particle size $R_0$. This value indicates the theoretical size-limit of particles that can be distinguished for a given sample and set-up. These were 52 nm and 73 nm in the third and fourth shifts, respectively.

**Background characterization**

The background scattering data were collected in the third shift, comprising 4 000 images taken with an average pulse energy of 1.135 mJ, as measured by the X-ray gas monitor detector (Maltezopoulos et al. 2019), and 120 000 images with an average pulse energy of 1.477 mJ in the fourth shift.

In addition to the instrument background, we measured the background including any contributions from the gas used for sample delivery itself, known as injection background, by using the frames classified as non-hits, as described above. We calculated the average injection background for each shift, except for the third shift when the detector was moved. As a result, we calculated two separate background profiles.

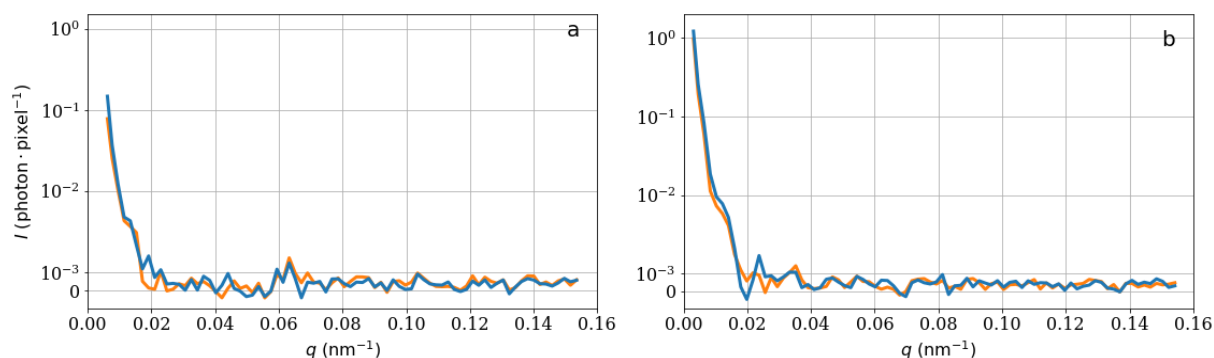

**Figure 3. Average background, in photons per detector pixel. a,b** Radially averaged background for the third and fourth shifts, respectively. The orange line is the instrument background and the blue line is the injection background. Note that the scale is linear below $10^{-3}$ photons per pixel.

The injection background, shown as a function of $q = \frac{2}{\lambda} \sin \theta$ (with θ half the scattering angle) in Fig. 3a,b, was averaged over 569 274 and 471 072 patterns with an average pulse energy of 1.276 and 1.539 mJ, respectively. The injection background barely exceeds the instrument background at low diffraction angles. The median background for all pixels of the detector was about $4 \times 10^{-4}$ photons per pixel in both shifts.

The background fades rapidly, reaching $10^{-3}$ photons per pixel from q > 0.02 nm$^{-1}$. The value of $10^{-3}$ photons per pixel is the limit of the statistical accuracy of background estimation, given the calibration of the AGIPD detector as available in this experiment (see Methods). At higher q, only stochastic fluctuations are observed.

**Variations in the position of diffraction pattern centers**

The position of the diffraction pattern centers varies from pulse to pulse since each particle collides with the X-ray beam at a random point relative to the beam axis (Loh et al, 2013). At these different interaction points, the beam has different phase shift values, that define the shift of the zero wavevector of the diffraction. The 2D histograms of the reconstructed centers of diffraction patterns scattered from spherical IrCl particles are shown in Fig. 4a,b. The diffraction pattern centers are given in horizontal ($\gamma_h$) and vertical ($\gamma_v$) angles of the beam deviation from the mean beam direction when measured from the interaction point.

The distribution during the third shift had an interquartile range (IQR) of 18 μrad along the horizontal axis and 20 μrad in the vertical direction. 90% of the diffraction pattern centers lie in the range of 50 and 59 μrad in the horizontal and vertical directions, respectively. During the fourth shift, the corresponding values of IQR were 18 and 22 μrad, and the corresponding ranges for 90% of the centers were 47 and 55 μrad. The fraction of centers inside the central pixel (see Fig. 4a,b, square shown in black dashed lines) is 91% and 94% for the third and fourth shifts, respectively.

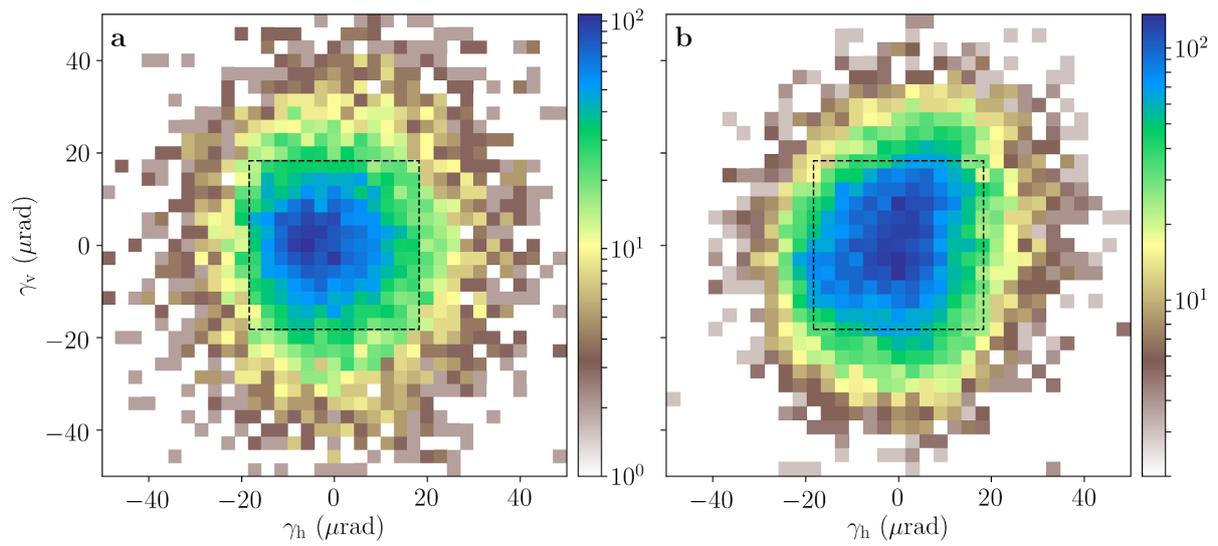

**Figure 4. Reconstructed positions of diffraction pattern centers.** 2D histograms of the distribution of the centers of diffraction patterns for the third (**a**) and fourth (**b**) shifts. The squares shown by black dashed lines indicate the edges of the detector pixel containing the center of the distribution. $\gamma_h$ and $\gamma_v$ are horizontal and vertical deviation from the mean beam direction when measured from the interaction point.

## Signal versus background

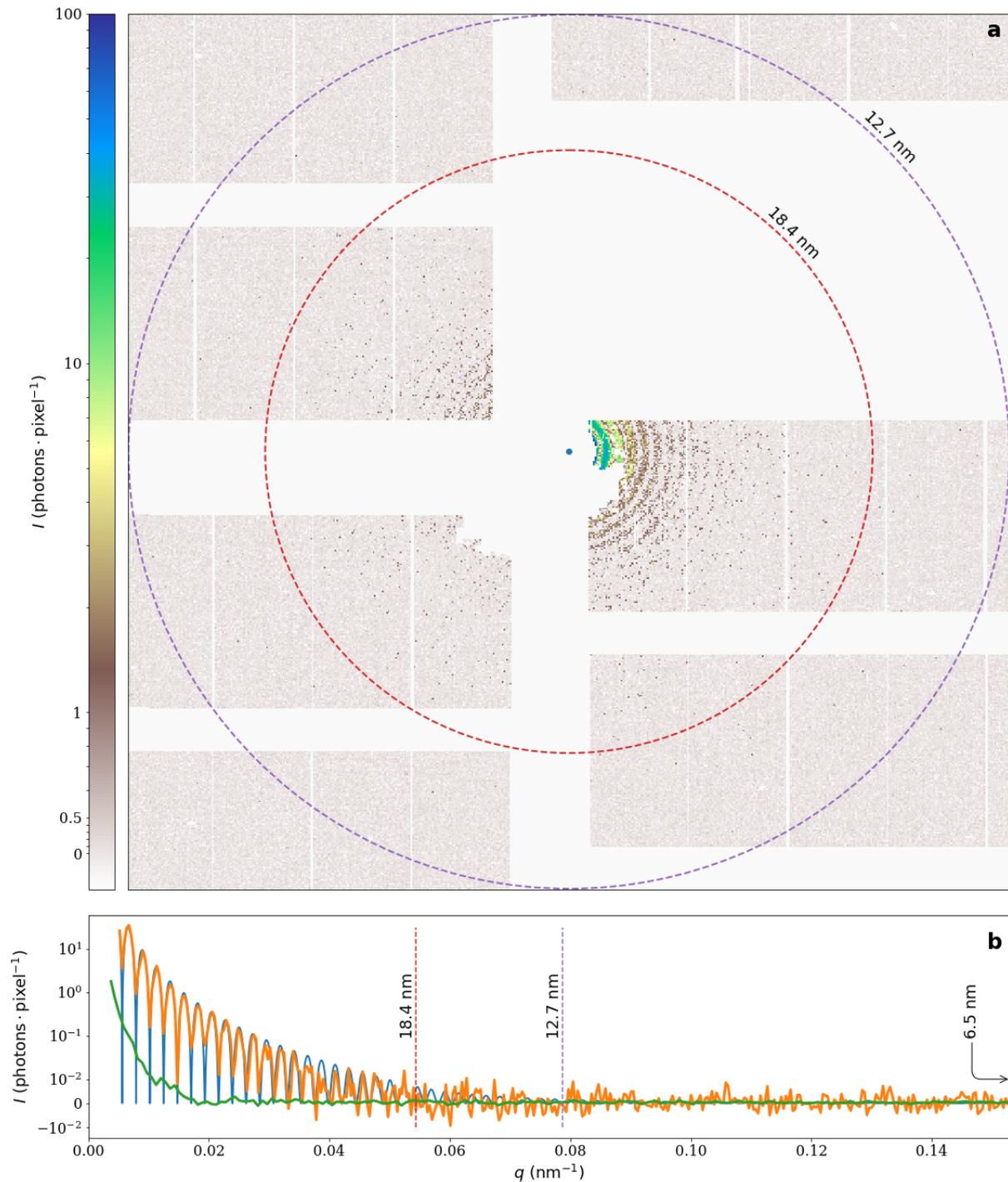

**Figure 5. a** Single strong diffraction pattern of an IrCl sphere of 439 nm in diameter, edge resolution is 12.7 nm. **b** comparison between the radially averaged scattering of the IrCl sphere (orange), fitted model (blue) and radially averaged background with injection (green). Note that the scale is linear below 10 photons per pixel. The red dashed lines (18.4 nm resolution) mark the angle at which the modeled scattering is stronger than the noise in a single frame; the purple dashed lines (12.7 nm resolution) mark the angle where the modeled scattering exceeds an average background; detector edge resolution is 6.5 nm.

The assembled and cropped diffraction pattern from a single hit of an IrCl particle is shown in Fig. 5a. The particle has an estimated diameter of 439 nm, which is close to the size of Mimivirus. The estimated incident photon fluence was $6.8 \times 10^8$ photons/µm².

The measured pattern corresponds to the spherical model at small diffraction angles (see Fig. 5b). At scattering vectors above 0.054 nm$^{-1}$ (red dashed line), the noise in one frame exceeds the amplitude of the spherical model, and fringes are not distinguishable, although the background when averaged across a large number of frames, is still an order of magnitude lower than the expected signal. The model approaches the injection background level at diffraction angles above 0.079 nm$^{-1}$ (purple dashed line).

**Filtering virus images by the particle size**

Scattering from Mimivirus particles was recorded in 154 runs, which produced a total of 4 million frames. A pixel where the signal was above one photon was considered to have detected photons, hereafter called a lit-pixel. Frames, where the number of lit-pixels was three standard deviations above the mean, were classified as hits and the rest as misses. This resulted in a set of 44 905 hit diffraction patterns, which were further processed.

The next step was to identify diffraction patterns produced by a single Mimivirus particle. In this work, we were only interested in single hit diffraction patterns as they can be immediately used to reconstruct the 3D Fourier space volume of the sample. To identify single hit diffraction patterns, we estimated the size of injected particles. A continuous wavelet transform (CWT)-based procedure was used, as described in the "Methods" section. The distribution of images by the diameter of the particle is presented in Fig. 6a.

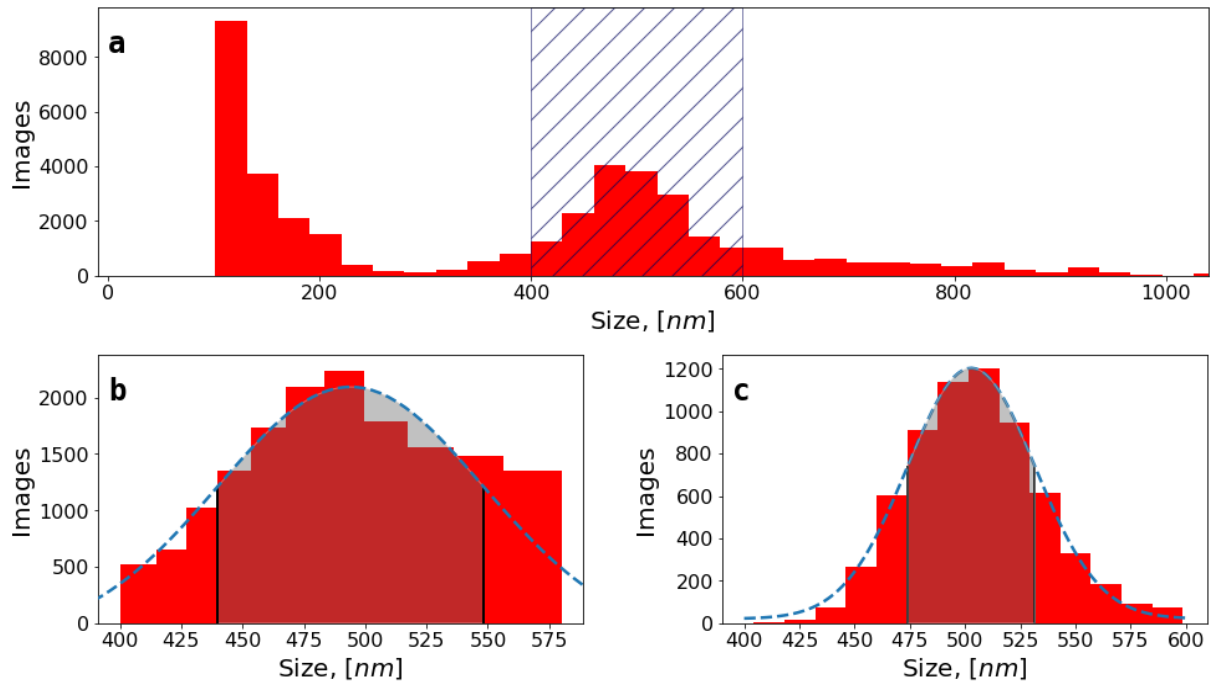

**Figure 6. Histogram of particle size distribution**. **a** - size distribution of all the particle diameters **b** - images with a particle diameter between 400 and 600 nm, **c** - images from the highlighted area in **b** with the recalculated diameters. Dashed blue line in **b,c** is the Gaussian fit. The highlighted region is mean plus or minus one standard deviation.

The particle diameter distribution (Fig. 6) is bimodal, with a maximum at the lower end of the detection range, which likely corresponds to aggregates of impurities (Bielecki et al. 2019), and another one at around 500 nm, which coincides with the diameter of Mimivirus particles measured by cryo-EM (Xiao et al. 2005). In the case of multiple hits, this size is significantly larger, and for non-virus particles, the size varies widely but is in general smaller than that of a Mimivirus.

In the distribution shown in Fig. 6a, we further selected the region of diameters from 400 nm to 600 nm (hatched area in Fig. 6a) and fitted it with a Gaussian distribution. We then discarded all images outside a one-sigma range and obtained a smaller subset of 11 308 diffraction patterns (see Fig. 6b). Relying on the fact that for these images we know the approximate particle size, we could use the last step of our CWT-based procedure (as described in "Methods") to recalculate that size more precisely (see Fig. 6c). We applied the one standard deviation criterion again, producing the final set of 4 335 images.

**Table 2.** Number of images and single hit ratio for different steps of filtering.

| total | identified | single hits | single hit ratio |
|---|---|---|---|
| 44 905 | 1 000 | 393 | 39±3% |
| 11 308 | 260 | 185 | 71±5% |
| 4 335 | 86 | 76 | 88±7% |

We randomly selected 1 000 images from the initial set of 44 905 hits, and manually identified single hits among them to estimate the efficiency of our filtering. In total, 393 images were marked as single hits. Out of the selected 1 000 images, 260 were part of the second set of 11 308 images with 185 of them having been marked as single hits. For the final set of 4 335 images, these numbers are 86 and 76, respectively. From these numbers, we can estimate the 95% confidence intervals for the ratio of single hits to all hits for each set (see Table 2), using the normal approximation. For the initial set this ratio is 39±3%, after the first step of filtering it becomes 71±5%, and in our final set of about 4 000 images 88±7% are single hits.

**Independence of the pulses within one train**

The small time interval between consecutive pulses of only around 1 microsecond in this experiment might have caused interference between adjacent pulses, e.g. due to the debris resulting from the preceding pulse. We investigated the distribution of incident photon fluences and particle sizes derived from spherical particles of IrCl for specific pulses within the trains (see Fig. 7a-d).

The distribution of particle sizes was different in the two shifts but remained stable over the pulses within a train. The incident photon fluences increased slightly throughout the first few pulses (up to 5 pulses), but then also remained stable up to the end of a train (Fig. 7c-d). This increase at the start of the pulse train agrees with the observed total pulse energy, as measured by the X-ray gas monitor detector of the instrument. The distributions of particle sizes for different pulses cannot be distinguished after taking into account the different incoming pulse energy, Fig. 7a-b. Therefore, we conclude that there was no correlation between pulse position in the train and particle size or incident fluence.

We also investigated the distribution of the number of patterns in one train, which could be fitted with the scattering model for spherical particles, hereafter called the number of fits. Details about when a fit was regarded as successful are described in the Methods. The frequency of fits is about the same for every pulse position in the train (see Fig. 7e).

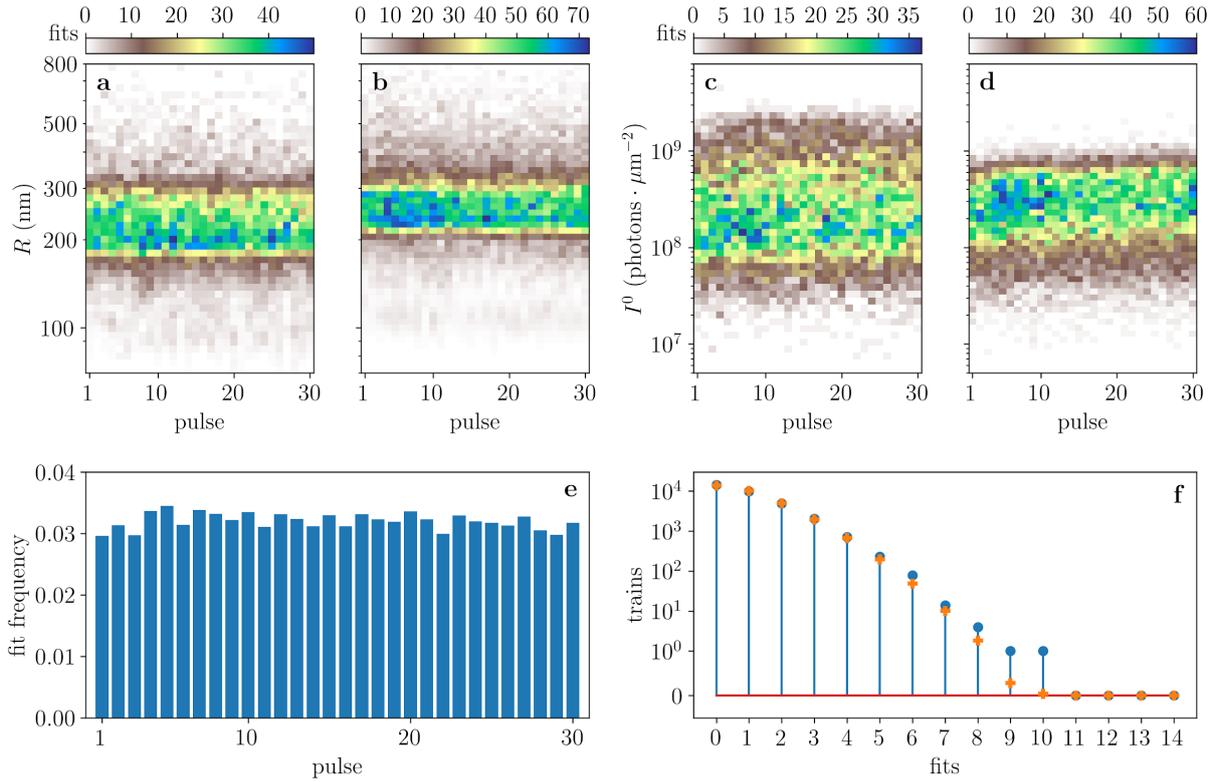

**Figure 7.** Characterization of IrCl hits across pulses within a train: **a,b** distribution of IrCl particle diameters for the third and fourth shifts, respectively; **c,d** distribution of incident photon fluences for the third and fourth shifts, respectively. **e** Fraction of shots that could be fitted as a function of the pulse position in the train. **f** Observed (blue) and expected (orange) histogram of the number of fits per train.

For independent pulses, the distribution of trains by the number of fits in them should follow a mixture of binomial distributions with the estimated probability of 'fit' events in individual runs equal to the fraction of successful fits in this train, the fit ratio,

$$G(k) = N \sum_i B\left(k, 30, \frac{M_i}{N}\right), B(k,n,p) = \frac{n!}{k!(n-k)!}p^k(1-p)^{n-k},$$

where $N$ is the number of frames in each run, $M_i$ is the number of fits in the run $i$, $k$ is the number of fits in a train, $i$ goes over runs.

A comparison of the expected distribution $G(k)$ and the observed distribution is presented in Fig. 7f. The two distributions agree very well, which is consistent with the independence of pulses in a train.

To additionally confirm the hypothesis of pulse independence, we computed the correlation coefficients of the derived spherical model parameters for all pairs of successive pulses and found no significant correlations between any of them.

## DISCUSSION

Coherent diffractive imaging requires a low-noise measurement of diffracted intensities from a sample. Even with the strong pulses available at XFELs, the number of diffracted photons from a single particle is relatively low due to the small scattering cross-section of X-rays. However, the high repetition rate of the EuXFEL allows the collection of very large datasets that can be used to improve the signal-to-noise ratio by averaging information from many diffraction patterns.

Background noise is an important determinant of the maximum resolution that can be achieved. The number of background photons per pixel in the first EuXFEL single particle experiment compares favorably with previous experiments at the CXI instrument of the LCLS (Daurer et al, 2017), although a quantitative comparison is difficult due to different experimental geometries. The detector is another critical component to achieve a low background, as it must be able to distinguish between electronic noise and real photons. The AGIPD detector demonstrated admirable performance, achieving a signal-to-noise ratio of 7 and being able to record data at an intra-bunch repetition rate of 1.1 MHz. Any instabilities in the instrument can lead to changes in the background making its removal much more difficult. Our measurements of the variation of the center of the diffraction patterns show an order of magnitude lower instability than similar measurements at the LCLS AMO instrument (Loh et al, 2013), and much smaller than one Shannon pixel.

The incident fluence on the sample is a key parameter for the success of single-particle imaging experiments. From the fits of the spherical patterns, we obtained a maximum beam fluence of about $2.8 \times 10^9$ photons/$\mu m^2$. This number is consistent with what one would expect from our experimental conditions; a 1 mJ pulse, 9 keV beam focused to a $15 \times 15$ $\mu m^2$ focal spot, resulting in $3.1 \times 10^9$ photons/$\mu m^2$, assuming perfect transmission. The relatively low maximum intensity, when compared to other XFEL experiments (Hantke et al. 2014, Daurer et al. 2017), is due to the initial larger temporary focus, which has since been upgraded.

The size estimates of the Mimivirus patterns show a peak around 500 nm, corresponding to the virus particles, and another one at the lower end of the detection range, below 200 nm. This second peak may be caused by contaminants in the solution which, combined with the large droplets created by the gas dynamic virtual nozzle (GDVN), can give rise to large aggregates (Bielecki et al., 2019). Using electrospray instead of GDVN for the formation of the aerosol is likely to eliminate this problem.

Statistical analysis shows that there are no correlations between pulses in the same train. The hit probability is also independent of the position of the pulse in the train or other hits in the same train. This clearly shows that any debris resulting from a hit leaves the interaction region before the next pulse arrives. It has been previously shown that for aerodynamic lenses, the main sample delivery instrument for X-ray single-particle imaging experiments, the particle speed increases with decreasing sample size (Hantke et al., 2018). This makes it likely that even at the maximum repetition rate of the EuXFEL, of 4.5 MHz, sub-100-nm particles should be able to vacate the interaction region in less than the minimum pulse spacing of 220 ns (Altarelli et al., 2014), making the maximum rate usable for most samples of interest.

## CONCLUSIONS

We presented an analysis of the first single-particle imaging experiment at the EuXFEL, performed when some of the functions planned for the SPB/SFX instrument were not yet available. However, the instrument proved to be very stable, and the measured background was low, which bodes well for future experiments. The measured photon flux in the interaction region matches what could be expected by taking into account the experimental conditions. The reduced focal spots achieved by the two Kirkpatrick-Baez-mirror pairs, which have since been installed at the instrument (Bean et al. 2016), should greatly improve the maximum flux, making future experiments with much smaller samples feasible. Measurements of smaller samples, however, would require changing injection from GDVN to electrospray, to avoid contamination due to the large droplets (Bogan et al. 2008; Bielecki et al. 2019; Uetrecht et al. 2019).

Despite the limitations in the available experimental parameters, in particular, focal spot and wavelength, we were able to conclusively demonstrate that it is possible to perform single-particle imaging at a megahertz repetition rate without any measurable difference between isolated and consecutive hits. This paves the path for high-repetition-rate and high data-rate single-particle imaging at XFELs.

## METHODS

### Sample preparation

An iridium(III) chloride hydrate (Sigma-Aldrich, purity 99.9%) solution at volume concentrations of 0.1% was used for the first 5 runs, and at a concentration 1% for the remaining runs. A solution of cesium iodide (Sigma-Aldrich, purity 99.9%) at a volume concentration of 1% was used for all respective runs. Melbourne and Mimivirus were both prepared following the protocol described in Okamoto et al. 2018, after which they were ultracentrifuged in sucrose gradient supplemented with 2.5% (v/v) glutaraldehyde to fixate them to fulfill the biosafety requirements of the EuXFEL. The fixed viruses were dialyzed five times in 250 mM ammonium acetate, pH 7.5 to remove the sucrose as completely as possible. Melbourne virus was used at a concentration of $10^{10}$ particles/ml in shift 4 and $2\times10^{10}$ particles/ml in the final shift. Mimivirus was used at a concentration of $10^{11}$ particles/ml in the first 11 runs of shift 3, at $3\times10^{11}$ particles/ml for the rest of shift 3 and the first 3 runs of shift 4, at $10^{12}$ particles/ml for the following 8 runs in shift 4, at $2\times10^{11}$ particles/ml for the next 42 runs in shift 4, at $10^{11}$ particles/ml for the final 47 runs in shift 4.

### Sample delivery

The samples were aerosolized using a gas dynamic virtual nozzle (GDVN) and focused on the interaction region as described in Hantke et al. 2018.

### Experimental set-up at the SPB/SFX instrument.

The data were collected at the SPB/SFX instrument of the EuXFEL in December 2017, under the proposal p2013. The accelerator produced 10 evenly spaced bunch trains per second with 30 X-ray pulses per bunch train at an intra-train repetition rate of 1.125 MHz, giving a separation between pulses of about 0.89 µs. The photon energy was 9.2 keV and the pulse energy, as measured by the gas monitor detector upstream, was around 1.5 mJ. The beam was focused by beryllium compound refractive lens (CRL) and the focus size was estimated to be 15 µm in diameter. The AGIPD 1M detector (Schwandt et al. 2013; Allahgholi et al. 2015; Allahgholi, A. et al 2019) was placed 5.465 m downstream from the interaction region. Online data analysis was done with Hummingbird (Daurer et al., 2016), through the Karabo bridge (Fangohr et al., 2017).

Beamline background on AGIPD was minimized using a three-slit collimation system as described in Kirby et al. 2013. Beam-defining 'power' slits made out of $B_4C$ were positioned close to the CRL on the downstream side. Further downstream, a set of anti-scattering slits, made from a tantalum-tungsten alloy, was used to clean up the stray light from the upstream optics. Finally, a set of germanium guard-slits was positioned far downstream, close to the sample position, in order to remove the secondary scattering produced by the anti-scattering slits. For all three slits, the gap was carefully adjusted, with micrometer accuracy, such that the slits received no direct beam while still maximizing the stray light reduction.

## Detector characterization

The AGIPD 1M detector (Schwandt et al. 2013; Allahgholi et al. 2015) contains 16 panels, each containing 64K pixels. The detector can record a signal from individual pulses in the bunch train, storing the data from each pulse into a separate memory cell on the chip. This results in variations of the detector response not only from one pixel to another but also between different memory cells of the same pixel.

The detector allows single-photon counting at 9.2 keV photon energy. We analyzed intensity histograms for each pixel and memory cell over all of the collected experimental data (see Fig. 8a). These histograms showed that the one-photon peak (located at $\mu_1$) was well separated from the zero-photon peak (located at baseline $\mu_0$). The baseline ($\mu_0$) and noise ($\sigma^0$) for each memory cell of each pixel were calculated as a mean and a standard deviation of the dark signal. The gain ($\mu_1 - \mu_0$) was determined from the difference between the first two peaks of pixel-cell intensity histogram.

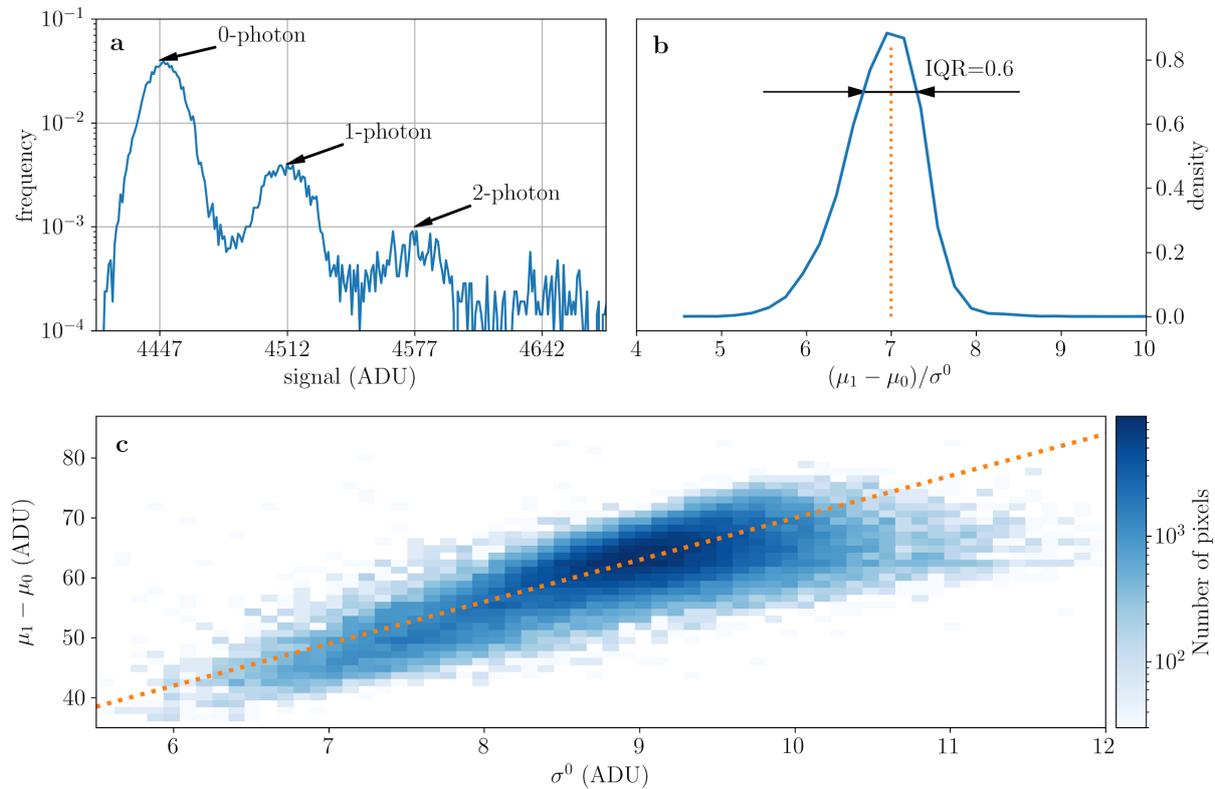

**Figure 8 | Detector gain characterization. a** Signal histogram for a single pixel. **b** Pixel-wise distribution of the ratio of gain to baseline noise. The orange dotted line shows the average SNR. **c** Pixel-wise distribution of the gain versus the noise.

A 2D histogram of the data by gain and noise is shown in Fig. 8c, and it shows a linear dependence between these parameters. The slope of the linear regression is equal to 7 and corresponds to the average signal-to-noise ratio (SNR) of the detector. The distribution of all SNR values is shown in Fig. 8b and has an interquartile range (IQR) of 0.6.

Only a small fraction of pixels had statistics sufficient to determine the 1-photon peak (at least about 100 events at the 1-photon peak). For the remaining pixels, to improve statistics we used histograms built using all memory cells of the same pixel. If the histogram-based grouping by the memory cells was still insufficient, we binned together blocks of 8 × 8 pixels to build a common histogram.

In cases when the single gain (g') parameter was determined for the group of memory cells or pixels by the combined histograms, the individual cell-pixel gain parameters were determined by multiplying g' on $\sigma_i/(\sum_i \sigma_i^2)^{1/2}$, where the summation is carried out over cell-pixels in the group.

Pixels with the noise ($\sigma$) or the baseline ($\mu_0$) values outside of a 3.5 standard deviations interval and with the gain ($\mu_1 - \mu_0$) outside of 4 standard deviations interval in the distributions of corresponding values over the detector panels were marked as bad pixels.

**Hit/non-hit images classification**

We used a lit-pixel counter (Hantke et al., 2014) to split frames into two classes: non-hits were frames with background scattering, and hits were frames with scattering from a sample.

In each frame, we calculated the number of lit pixels that record a signal of more than 45 analog-to-digital units above the baseline (~0.7 of the one-photon signal). For each run, the histogram of lit-pixel counts was fitted with a Gaussian function. The value equal to 2.5 standard deviations above the mean of the fitted Gaussian was set as a threshold for the hits in this particular run. Frames with the number of lit pixels below the threshold were classified as non-hits. If we had a true Gaussian distribution of lit-pixels in the set of frames only with background scattering, then we would expect about 150 (~0.5%) false positive hits per run using this value of the threshold.

**Model of scattering from spheres**

The scattered intensity from a sphere of diameter $R$, placed in the beam with incident photon fluence $I_0$ at the scattering vector $q$ is given by

$$I(q, R, I^0) = I^0 \left(r_e \frac{\pi R^3}{6} n\right)^2 \Delta\Phi \left[3 \frac{j_1(\pi qR)}{\pi qR}\right]^2,$$

where $n$ is the density of electrons, $r_e$ is the classical electron radius, $\Delta\Phi$ is the solid angle and $j_1$ is the spherical Bessel function of the first kind.

The length of the scattering vector related to the $i$-th pixel with coordinates $(x_i, y_i)$ on the detector at the distance $L$ from the scattering point is

$$q_i = \frac{2}{\lambda} \sin \theta_i = \frac{\sqrt{2 - 2c_i}}{\lambda}, c_i = \cos 2\theta_i = \frac{L}{\sqrt{L^2 + r_i^2}}, r_i = \sqrt{(x_i - x)^2 + (y_i - y)^2},$$

where $x, y$ are the coordinates of the diffraction pattern center, $\lambda$ is the wavelength, $2\theta_i$ is the angle between the beam direction and the direction to the pixel $i$.

The solid angle of $i$-th pixel is

$$\Delta\Phi_i = \frac{A}{L^2} c_i^3,$$

where $A$ is an area of a pixel.

The measured diffraction $v_i$ at pixel $i$ is a result of the combination of Poisson and Gaussian statistics

$$v_i = P(I_i + b_i) + N(0, \sigma_i^2),$$

where $\sigma_i$ is the instrumental error at the pixel $i$, estimated by the processing of the dark run, and $b_0$ is the averaged background scattering.

One diffraction pattern consists of $N$ pixels with successfully measured diffraction

$$X = \{x_i, y_i, v_i, \sigma_i, b_i\}, i = 1 \ldots N.$$

**Fitting the model of scattering from spheres to experimental pattern**

The following procedure was used for model-based interpretation of the experimental diffraction pattern $X$:
1. Finding a rough estimate of the center $(x, y)$ of the diffraction pattern averaged over several strongest patterns using the Hough transform (Rosenfeld 1969; Hough 1962).
2. Finding a rough estimate of the diameter $R$ of the particle and the incident photon fluence $I^0$ by a least-squares fit of the scattering from the spherical model to the measured radially averaged diffraction intensity.
3. Selecting the interpretable images according to $\chi^2$ value of the fit.
4. Refining all parameters $(x, y, R, I^0)$ using maximum likelihood given the measured intensities $(v_i)$. In contrast to step 2 here, we also refine the center of the diffraction pattern.

**Refinement of parameters with likelihood maximization**

Here, we approximate the Poisson distribution with the Normal distribution. Then the likelihood may be written as

$$\mathcal{L}(\theta|X) = \prod_{i=1}^{N} \frac{1}{\sqrt{2\pi(I_i+\sigma_i^2)}} \exp\left(-\frac{(I_i+b_i-v_i)^2}{2(I_i+\sigma_i^2)}\right).$$

Take a logarithm

$$l(\theta|X) = -\frac{1}{N}\log \mathcal{L}(\theta|X) = \frac{\log 2\pi}{2} + \frac{1}{2N}\sum_{i=1}^{N}\log(I_i+\sigma_i^2) + \frac{1}{2N}\sum_{i=1}^{N}\frac{(I_i+b_i-v_i)^2}{(I_i+\sigma_i^2)}.$$

The optimal parameters correspond to the minimum of $l$

$$\theta = (R, I^0, x, y) = \arg\min\, l(\theta|X).$$

The goodness of fit was estimated as

$$\chi^2 = \frac{1}{N}\sum_{i=1}^{N}\frac{(I_i-v_i)^2}{I_i+\sigma_i^2}.$$

The fitting was regarded as successful if the first- and the second-order optimality conditions were met and the goodness of fit ($\chi^2$) was less than a predefined tolerance

$$\left\|\frac{\partial \theta}{\partial X}\right\| < \varepsilon,\, H = \frac{\partial^2 l}{\partial \theta \partial \theta'}\text{ is positive defined}, \chi^2 < \zeta,$$

where $\varepsilon$ and $\zeta$ are predefined tolerance. We used $\varepsilon = 10^{-6}$ and $\zeta = 1.1$

**Fast determination of particle size by the CWT**

To estimate the size of the scattering particle for each diffraction pattern we used the spherical particle model. A centered diffraction pattern is converted to its radial average which is then compared to the diffraction pattern of a uniform sphere. To account for an unknown background signal present in experimental data, the experimental and theoretical spherical diffraction functions were only compared at the positions of their maxima.

To find peaks in noisy experimental radial average, we used a CWT-based peak detection algorithm (Du et al. 2006). We used, scaled and translated the second peak of the spherical form factor as our wavelet, which has produced better results than the commonly used Ricker wavelet.

To estimate the diameter of the particle, we used three passes of this CWT procedure. The first pass was tuned to identify images for which the diameter was too small (less than 300 nm); these images were discarded. The second pass was used to estimate the diameter of larger particles with a diameter between 300 and 800 nm. In both cases, we estimated the diameter using the average distance between neighboring maxima, relying on the fact that for spherical form factor this distance is very close to $\pi/r$.

The third pass was used to refine the initially determined approximate value of the particle diameter. We used the positions of the first three peaks in the spherical scattering function to refine the particle size using least-squares minimization.

$$\frac{x_i}{r} + s,$$

Where $X_i$ is a position of i-th order maximum of spherical form factor with 1 nm radius and $s$ is arbitrary constant shift introduced to account for imprecise determination of the center of the diffraction and for the fact that experimental particles are not perfectly spherical. In this way, in addition to the particle diameter, we obtain two more values - the shift of the beam center and the mean square error of the fit. Both these values are used to estimate the reliability of the obtained parameters.

## Data Availability

Data are available from the corresponding author upon reasonable request.

## Acknowledgements

We acknowledge European XFEL in Schenefeld, Germany, for provision of X-ray free-electron laser beamtime at Scientific Instrument SPB/SFX and would like to thank the instrument group and facility staff for their assistance. The results of the work were obtained using Maxwell computational resources operated at Deutsches Elektronen-Synchrotron (DESY), Hamburg, Germany, and computational resources of MCC NRC "Kurchatov Institute". This research used resources of the National Synchrotron Light Source II, a U.S. Department of Energy (DOE) Office of Science User Facility operated for the DOE Office of Science by Brookhaven National Laboratory under Contract No. DE-SC0012704. We acknowledge the support of funding from: the Swedish Foundation for International Cooperation in Research and Higher Education (STINT); Helmholtz Associations Initiative and Networking Fund and the Russian Science Foundation grant HRSF-0002/18-41-0600; European Research Council, "Frontiers in Attosecond X-ray Science: Imaging and Spectroscopy (AXSIS)", ERC-2013-SyG 609920; Fellowship from the Joachim Herz Stiftung; Singapore National Research Foundation Grant number NRF-CRP16-2015-05; Ministry of Education, Science, Research and Sport of the Slovak Republic and by grant APVV-18-0104; the project CZ.02.1.01/0.0/0.0/16_019/0000789 (ADONIS) from European Regional Development Fund, Chalmers Area of Advance; Material Science and the Ministry of Education, Youth and Sports as part of targeted support from the National Programme of Sustainability II; US National Science Foundation (NSF) Science and Technology Center BioXFEL Award 1231306; Helmholtz Initiative and Networking Fund through the Young Investigators Group Program and Deutsche Forschungsgemeinschaft, project B03/SFB755; VR starting grant (2018-03387); FORMAS future research leader (2018-00421); KVA Biosciences 2018 (BS2018-0053); NSF 1231306; German Ministry for Education and Research, BMBF (grant No. 05K2016 - Visavix); the Heinrich Pette Institute, Leibniz Institute for Experimental Virology is supported by the Free and Hanseatic City of Hamburg and the Federal Ministry of Health; NSF STC BioXFEL grant 1231306; The National Research Foundation (NRF) of Korea (Grant No. 2017K1A3A7A09016380); the Röntgen-Ångström Cluster; the Swedish Research Council; the Swedish Foundation for Strategic Research. We thank Arwen Pearson for critical reading of the manuscript.

## Author Contributions

K.G., J.B., A.B., H.N.C., T.E., A.O., C.U., I.V., A.M., and F.R.N.C.M. conceived and designed the experiments. K.O., H.K.N.R., B.G.H. and prepared samples. K.O., H.K.N.R., B.G.H., O.K., Y.K., P.L.X, and R.S. characterized the samples. J.B., J.A., S.Bari, L.F., D.A.H., Y.K., R.A.K., O.K., K.M., A.V.R. and M.T. developed and operated the sample delivery equipment. E.S., S.Z., K.A., I.B., A.B., S.Bobkov, G.C., B.J.D., T.E., O.G., L.G., A.H., V.I., D.L., P.S., J.A.S., A.T. and F.R.N.C.M contributed to data processing and analysis. K.G., J.B., R.B., K. D., Y.K., H.K., R.L., M.M., A.R., T.S., M.S., S.S., B.W. and A.M developed and operated the SPB/ SFX instrument at EuXFEL. N.R., J.S. and S.H. operated the detector at EuXFEL. A.S. and C.X. developed EuXFEL DAQ and controls. The manuscript was written by E.S., S.Z., I.V., V.S.L. and F.R.N.C.M. with input from all authors.

**Competing interests:** The authors declare no competing interests.